\documentclass[12pt]{article}

\usepackage{epsf}

\newcommand{\bmat}{\left(\begin{array}}
\newcommand{\emat}{\end{array}\right)}
\def\NPB{Nucl. Phys. B}

\def\yzero{\smash{\hbox{$y\kern-4pt\raise1pt\hbox{${}^\circ$}$}}}

\def\a{\alpha}

\def\n{{\tilde n}}
\def\m{\tilde m}

\def\beq{\begin{equation}}
\def\eeq{\end{equation}}
\def\beqa{\begin{eqnarray}}
\def\eeqa{\end{eqnarray}}
\def\ba{\begin{array}}
\def\ea{\end{array}}

\def\-{\hphantom{-}}

\def\s2{\frac{1}{\sqrt2}}

\def\beq{\begin{equation}}
\def\eeq{\end{equation}}
\def\beqa{\begin{eqnarray}}
\def\eeqa{\end{eqnarray}}

\def\IF{\relax{\rm I\kern-.18em F}}
\def\II{\relax{\rm I\kern-.18em I}}
\def\IP{\relax{\rm I\kern-.18em P}}
\def\IC{\relax\hbox{\kern.25em$\inbar\kern-.3em{\rm C}$}}
\def\IR{\relax{\rm I\kern-.18em R}}

\def\cp{{\cal P}}

\def\Dsl{\,\raise.15ex\hbox{/}\mkern-13.5mu D} 
\def\IZ{Z\kern-.4em  Z}

 \def\cp#1{\relax\ifmmode {\IP\kern-2pt{}_{#1}}\else $\IP\kern-2pt{}_{#1}$\=fi}
\newcommand{\drawsquare}[2]{\hbox{%
\rule{#2pt}{#1pt}\hskip-#2pt
\rule{#1pt}{#2pt}\hskip-#1pt
\rule[#1pt]{#1pt}{#2pt}}\rule[#1pt]{#2pt}{#2pt}\hskip-#2pt
\rule{#2pt}{#1pt}}

\newcommand{\fund}{\raisebox{-.5pt}{\drawsquare{6.5}{0.4}}}
\newcommand{\Ysymm}{\raisebox{-.5pt}{\drawsquare{6.5}{0.4}}\hskip-0.4pt%
        \raisebox{-.5pt}{\drawsquare{6.5}{0.4}}}
\newcommand{\Yasymm}{\raisebox{-3.5pt}{\drawsquare{6.5}{0.4}}\hskip-6.9pt%
        \raisebox{3pt}{\drawsquare{6.5}{0.4}}}
\newcommand{\antifund}{\overline{\fund}}

\catcode`\@=11
\newdimen\@rotdimen
\newbox\@rotbox

\def\@vspec#1{\special{ps:#1}}
\def\@rotstart#1{\@vspec{gsave currentpoint currentpoint translate
   #1 neg exch neg exch translate}}
\def\@rotfinish{\@vspec{currentpoint grestore moveto}}
\def\@rotr#1{\@rotdimen=\ht#1\advance\@rotdimen by\dp#1%
   \hbox to\@rotdimen{\hskip\ht#1\vbox to\wd#1{\@rotstart{90 rotate}%
   \box#1\vss}\hss}\@rotfinish}
\def\@rotl#1{\@rotdimen=\ht#1\advance\@rotdimen by\dp#1%
   \hbox to\@rotdimen{\vbox to\wd#1{\vskip\wd#1\@rotstart{270 rotate}%
   \box#1\vss}\hss}\@rotfinish}%
\def\@rotu#1{\@rotdimen=\ht#1\advance\@rotdimen by\dp#1%
   \hbox to\wd#1{\hskip\wd#1\vbox to\@rotdimen{\vskip\@rotdimen
   \@rotstart{-1 dup scale}\box#1\vss}\hss}\@rotfinish}%
\def\@rotf#1{\hbox to\wd#1{\hskip\wd#1\@rotstart{-1 1 scale}%
   \box#1\hss}\@rotfinish}%
\def\rotate{\@ifnextchar[{\@rotate}{\@rotate[l]}}
\def\@rotate[#1]#2{\setbox\@rotbox=\hbox{#2}\@nameuse{@rot#1}\@rotbox}

\catcode`\@=12

\begin{document}

\makeatletter \@addtoreset{equation}{section} \makeatother
\renewcommand{\theequation}{\thesection.\arabic{equation}}
\pagestyle{empty}
\pagestyle{empty}
\rightline{UOA/NPPS-08/04}
\vspace{1.0cm}
\setcounter{footnote}{0}


\begin{center}
\Large{ 
{\bf  
Standard Model Compactifications of $IIA$ $Z_3 \times Z_3$ Orientifolds
from Intersecting D6-branes
}}
\\[4mm]
{\large{ Christos Kokorelis 
}
\\[1mm]}
\normalsize{\em   Institute of Nuclear Physics, N.C.S.R. Demokritos, 
GR-15310, Athens, Greece\\
and\\
   Nuclear and Particle Physics Sector, Univ. of Athens,
GR-15771 Athens, Greece
} 
\end{center}                  
\vspace{1.0mm}
\begin{center}
{\small \bf Abstract}
\end{center}

\begin{center}
{\small  ABSTRACT}
\end{center}
We discuss the construction of chiral four dimensional    
${\bf T^6/(Z_3 \times Z_3)}$ orientifold compactifications of 
IIA theory, using D6-branes intersecting at
angles and not aligned with the orientifold O6 planes.  
Cancellation of mixed 
U(1) anomalies requires the presence of a generalized Green-Schwarz mechanism 
mediated by RR partners of closed string untwisted moduli.  
In this respect we describe the appearance of 
 three quark and lepton family $SU(3)_C  \times SU(2)_L \times U(1)_Y$ 
non-supersymmetric orientifold 
models with only the massless spectrum of the SM at low energy that can have either 
no exotics present and three families of $\nu_R$'s (A$^{\prime}$-model class) or 
the massless fermion spectrum of the N=1 SM with a small number of
massive non-chiral colour exotics and in one case with extra families 
of $\nu_R$'s (B$^{\prime}$-model class).
Moreover we discuss the construction of SU(5), flipped SU(5) and Pati-Salam 
$SU(4)_c \times SU(2)_L \times SU(2)_R$ GUTS - the latter also
derived from adjoint breaking - with only the SM at low energy. 
Some phenomenological features of these models are also briefly discussed.
All models are constructed with the Weinberg angle to be 3/8 at the string scale.

\newpage

\setcounter{page}{1} \pagestyle{plain}
\renewcommand{\thefootnote}{\arabic{footnote}}
\setcounter{footnote}{0}

\section{\large Introduction}

Four dimensional perturbative (4D) chiral compactifications (CS) from string theory 
have a long 
history. The first perturbative chiral models started with 
the N=1 4D 
closed string compactifications
of the heterotic string - on either Calabi-Yau manifolds \cite{chiral1}, 
orbifold \cite{chiral2}, self-dual lattices \cite{chiral3}, 
Gepner type \cite{chiral4}, fermionic constructions \cite{chiral5} or orbifolds with 
Wilson lines \cite{wil}. In these 
vacua the string scale is of the order of $10^{18}$ GeV, the gauge 
couplings unify at the same scale and as a
consequence - of the high scale - proton is stable. As the string scale is high 
N=1 susy models are favored phenomenologically, also offering simultaneously 
a solution to the gauge hierarchy problem. 
However, these N=1 vacua face severe problems related to the breaking of 
supersymmetry  - creating a non-zero cosmological constant - 
and also lack satisfactory moduli fixing mechanisms.

On the other hand compact type II orientifold \cite{openrev} compactifications 
offer a  
window into perturbative physics - that has been used in recent 
years to study supersymmetric and 
non-supersymmetric D-brane models (See \cite{rev} for reviews)- as in type I
compactifications the gauge hierarchy problem could be solved 
by the existence of dimensions transverse to the space that the D6-branes 
wrap \cite{anto}.  
In particular, stringy D-brane models derived from orientifolds of type II 
string compactifications that include  
D6-branes intersecting at angles (IBs)
[13-46] and wrapping three cycles in the six dimensional internal space,  
provide a consistent string
framework that determines all the physical quantities in terms of the brane 
angles and their wrappings.     
In these constructions chiral fermions get localized in the intersections between 
branes \cite{dou}.

Various compact chiral IIA orientifold N=1 supersymmetric constructions with 
intersecting D6-branes have been produced including 
orbifolds of $T^6/Z_2 \times Z_2 $ \cite{cve},  $T^6/ Z_4 $ \cite{bluo1},  
$T^6/ Z_4 \times Z_2 $ \cite{gabi},  $T^6/ Z_6 $ \cite{ottho}.  However, the effect of 
N=1 supersymmetry seriously affects the spectrum, as in all cases the spectra are 
semi-realistic with the 
N=1 SM accompanied by either massless chiral \cite{cve, gabi} (class ${\tilde A}$) or 
massless non-chiral 
exotics (class ${\tilde B}$) \cite{bluo1, ottho}.     
We also note that chiral models could be produced from orientifolds of Gepner 
models \cite{gepner} 
using 
techniques borrowed from intersecting branes. In this 
case the models of class ${\tilde B}$ have been produced \cite{she}.

On the other hand, the first attempts in a string theory context resulting in 
non-susy 
chiral constructions 
with intersecting D6-branes wrapping 3-cycles [ In the T-dual language these models 
correspond to models with magnetic deformations \cite{ang, pra}]
 have been carried out in 
compactifications of type IIA 
on either tori  \cite{ibaba} or
toroidal orientifolds $ IIA/(T^6/\Omega R)$ \cite{lust1}, the latter using the T-dual 
picture with D9-branes and background fluxes turned on, following the work 
of \cite{bachas} on the 
gauge theory aspects of magnetic fluxes. 
Three generation non-supersymmetric models with no extra exotics 
have been found in 
toroidal 
orientifolds (TO)\cite{lust1} or $Z_3$ orientifolds \cite{lust3}.  
Indeed using the constructions \cite{lust1} it has become possible to 
derive - using only bifundamental representations -  non-supersymmetric 
vacua that possess {\em just the observed Standard 
Model} (SM) spectrum with right handed neutrinos and gauge interactions at low energies 
\cite{louis2, kokos5, kokos6} and also Pati-Salam vacua with the same property \cite{kokos1, kokos2}.
[In \cite{lust3}, non-supersymmetric Standard model vacua have been also derived from $Z_3$ 
orientifolds, but 
due to existence of antisymmetrics in the spectrum there were no mass terms for the up-quarks.]
Important results of these TO constructions \cite{kokos2} include a) {\em The fixing of all 
complex 
structure moduli using N=1 supersymmetry in open string sectors in toroidal 
orientifolds \footnote{e.g. see eqn. (4.37) in hep-th/0203187} } 
and the fact that b) 
{\em N=1 supersymmetry conditions that introduce supersymmetric sectors - 
in toroidal 
orientifolds - solve the condition that 
the hypercharge survives massless 
the Green-Schwarz anomaly cancellation  \footnote{The latter mechanism acts as 
a mass generation mechanism 
to U(1)'s that have a non-zero coupling to RR fields.} mechanism}.
In all D-brane models coming from intersecting branes, 
the SM is accompanied by the simultaneous existence of right handed
neutrinos; necessary for RR tadpole cancellation [other proposals in   
D-brane model building but not based to a particular string construction can be seen in
 \cite{antotoma}, \cite{be}].

The purpose of this work, is twofold. Firstly to discuss the main features of the 
${\bf Z_3 \times Z_3}$ orientifolds and to also 
show that it is not only possible to produce
(3-stack) non-supersymmetric models
with only the SM at low energy - reproducing the SM fermion spectra of \cite{lust3} - 
but also derive new non-supersymmetric vacua which localize the fermion 
spectrum of the N=1 
SM (3- and 5-stack) and extra massive exotics which subsequently break to the SM. These SMs 
exhibit partial gauge unification of the strong and weak 
gauge couplings with a Weinberg angle $sin^2 \theta = 3/8$, as it has been also shown in the 
split susy scenario \cite{split0}[see also section 8 of \cite{split01} for different models - distinguished by the different 
electric and magnetic wrappings that solve the RR tadpoles and intersection numbers - with the same 
spectrum]. 
Secondly,  
to show that GUT models could be constructed. We present  
explicit examples with two stack flipped SU(5) GUTs and three stack 
Pati-Salam $SU(4)_c \times SU(2)_L \times SU(2)_R$ GUTS.

The paper is organized as follows. In section two we discuss the formalism of the chiral 
constructions on the 
$Z_3 \times Z_3$ orientifolds. We describe explicitly the derivation of the constraints coming
from RR tadpole cancellation at the string scale. 
RR tadpoles constitute an important constraint in string model
building as their presence is equivalent to the constraints coming from the 
cancellation of cubic gauge 
anomalies in the low energy effective theory. 
We also describe the spectrum  
rules for the $Z_3 \times Z_3$ orientifold vacua that are based on the AAA tori 
lattices. The explicit form of the effective wrappings is seen in appendix A.  
In addition spectrum rules and RR tadpoles for different $Z_3 \times Z_3$ 
orientifold vacua - for which we will not present explicit non-susy or 
N=1 supersymmetric models 
in this work - 
are presented in appendix B. 
   In section 3 we discuss the structure of the U(1) anomaly 
cancellation necessary for the cancellation of U(1)-mixed gauge/gravitational anomalies.
The rest of the sections is devoted to the study of non-supersymmetric models which in most of the
cases exhibit partial unification of their gauge couplings, that is they 
unify - at the string scale - the strong $SU(3)_c$ and the weak $SU(2)$ gauge couplings with the
Weinberg angle to be $sin^2 \theta = 3/8$ as in the successful SU(5) GUT prediction.      
In section 4 we present three stack non-susy models which break to only the SM at low energy without 
any massless exotics present. These models have been also found before 
in the 
$Z_3$ orientifolds of \cite{lust3}. In section 5, we derive other three stack three generation 
non-supersymmetric  models with the Weinberg angle $sin^2 \theta = 3/8$ which 
localize the massless fermion spectrum of the N=1 supersymmetric SM 
(with extra generations of $\nu_R$'s) in the presence of one pair of non-chiral exotics.
Eventually, the extra beyond the SM massless fermions and the extra exotics 
become massive leaving only the SM at low 
energy 
[In our companion paper \cite{split01} we presented SMs which
generate models with the same chiral spectrum but for different wrapping numbers]. \newline
In section 6 -  even though the present models do not exhibit a supersymmetric 
spectrum -
we discuss the split susy scenario which was proposed as an 
alternative signal for LHC. 
Hence a comparison with split susy models (SSS) that appear recently in the literature is also
performed as it appears that even though the Z3 x Z3 models are not SSS, they do possess 
many of their properties. As an application, we discuss
five stack vacua with the spectrum of the N=1 SM and massive exotics. The models break to 
the SM at low energy as the Higgsinos form massive Dirac pairs.
In section 7, we discuss the construction of flipped SU(5) GUTS with the SM at low 
energy. Models with the same fermion spectrum have been produced in \cite{axeflo} 
based on the $Z_3$ orientifolds \cite{lust3}. 
In sections 8, 9 we discuss the construction of SU(5) and Pati-Salam 
$SU(4)_L \times SU(2)_L \times SU(2)_R$ (with adjoint breaking) non-supersymmetric 
three family GUTS respectively.
Section 10 contains our conclusions.

\section{\large RR Tadpoles and Spectrum rules for ${\bf Z_3 \times Z_3}$ 
Orientifolds} 

In this section we will describe the spectrum rules and the 
constraints on the parameters (wrappings) that have to be imposed on model building 
attempts on these  
 constructions. The existence of these rules is independent of the      
presence of supersymmetry - in the open string sector -  where the chiral matter of 
the Standard model gets localized.

\subsection{\small The background and the RR tadpole cancellation}

Our constructions are derived from IIA theory compactified on 
a six dimensional torus modded out by the orbifold action     
${\bf (Z_3 \times Z_3)}$, where the latter symmetry is generated by the twist
generators \footnote{where $\a = e^{\frac{2\pi i }{3} }$,}
\beqa   
\theta : (z_1, z_2, z_3) \rightarrow (\a z_1, \a^{-1} z_2, z_3) ,\nonumber\\ 
\omega :(z_1, z_2, z_3) \rightarrow ( z_1, \a z_2, \a^{-1}z_3) ,
\label{asdq1}
\eeqa
and where $\theta $, $\omega$ gets associated to the action of the 
twists 
$\upsilon =  \frac{1}{3}(1, -1, 0)$, 
$ u = \frac{1}{3}(0, 1, -1)$.  Here, $z^i = x^{i+3}+ i x^{i+5}$, 
$i=1,2,3$ are the complex coordinates on   
$T^6$, which we consider as being factorizable for simplicity reasons, namely
$T^6 \equiv T^2 \otimes T^2 \otimes T^2$.
In addition to the orbifold action, 
IIA is also modded out by the orientifold 
action $\Omega R$,  that 
combines the worldsheet parity  $\Omega $ and the antiholomorphic involution 
\beq
R : z^i \rightarrow {\bar z}^i  \    .
\label{loc}
\eeq
The orbifold action has to act crystallographically on the 
lattice.
 For this reason, we will let the complex structure on all 
three $T^2$ tori to be fixed at
\beq
U^I_A = \frac{1}{2} - i\frac{\sqrt{3}}{2} \ , 
\eeq
We associate the presence of the above complex structure with the A-torus.  
The A-torus - will be used in the main body of the paper in which we discuss model building 
using the AAA tori - is a modified version 
of the A-torus that appear in \cite{lust3}, as we have chosen the 
lattice vectors to be $e_1 =(1,0)$, $e_2=(-1/2, \sqrt{3}/2)$.
The 
orbifold action (\ref{asdq1}) preserves N=2 SUSY in four dimensions 
and thus the orbifold  
describes the singular limit of a Calabi-Yau threefold. 
Using the cohomology of the internal orbifold space we get from 
the Hodge numbers describing this threefold \footnote{This class 
of orbifolds corresponds to models without discrete torsion.}   
that in the closed string sector are $h^{11} = 84$, $h^{21} = 0$, where
three K\"ahler moduli come from the untwisted sector and the rest from the 
twisted sectors. 
As the numbers of the 
independent three cycles is $b_3 = 2 + 2 h^{21}$ all the independent cycles
are inherited from the six dimensional toroidal space. There are no 
exceptional cycles involved and hence we will need only toroidal cycles,
 as in the orientifolds of $T^6/Z_3$ \cite{lust3}, 
$T^6/(Z_2 \times Z_2)$  \cite{cve}.

\begin{figure}
\centering
\epsfxsize=6in
\hspace*{0in}\vspace*{.2in}
\epsffile{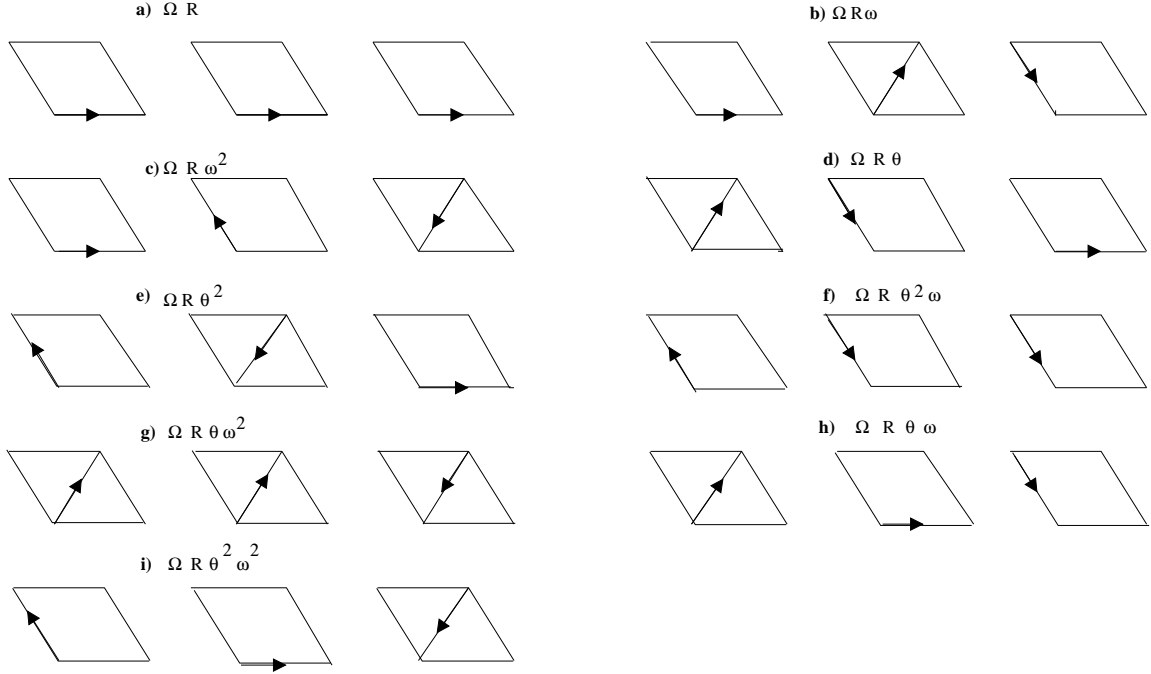}
\caption{\small
O6-planes in the orientifold of $T^6/(Z_3 \times Z_3)$}
\label{figo1}
\end{figure}

The orientifold projection breaks the bulk supersymmetry to N=1 and introduces 
orientifold O6-planes locked at the fixed locus of the antiholomorphic involution (\ref{loc}).
The O6-planes carry negative RR charge whose presence - introduces an 
inconsistency into the theory as the partition function diverges - 
can be cancelled by the
introduction of D6-branes intersecting at angles with the O6-planes.
The models that are derived are chiral as the D6-branes are not parallel to the
orientifold planes [Non-chiral 4D models on $Z_3 \times Z_3$ orientifolds of type IIA have been 
considered in the past, in 
the presence of intersecting D-branes parallel to the O6-planes in \cite{twisted2} and in a
different content in \cite{prala}].  
The D6-branes are assumed to be 
wrapping 3-cycles along the toroidal space, 
with the 1-cycles being described
by the (`electric'-`magnetic') numbers respectively $(n^i_a, m^i_a)$, 
indicating wrapping along i-th cycle of the $T^2$ tori.
There are four possible tori 
choices, allowing for a consistent twist action within the $Z_3 \times Z_3$ orientifold
that is the AAA, AAB, ABB, BBB one's. In this work we will exhibit model building attempts 
which are based on the AAA tori in the main body of this paper. In appendix B, we will 
derive the spectrum rules and RR tadpoles for the vacua that accommodate the AAB, BBB tori using
the A, B torus choices made in  \cite{lust3}.
 Under the orbifold and orientifold action the branes are organized
into orbits of length nine. These orbits are characterized by pairs of wrapping 
numbers  $(n^i, m^i)$. Especially for the AAA torus that we will treat in detail in this work, 
these orbits are described by 
\beqa
\left( \ba{c} n^i\\ m^i
\ea \right)  \  \stackrel{Z_3}{\Longrightarrow} 
& 
\left(\ba{c}- m^i\\ n^i - m^i
\ea \right)   \ \stackrel{Z_3}{\Longrightarrow}   &
\left(\ba{c}-n^i+ m^i\\ - n^i \ea \right)\nonumber\\
 \Omega R\Downarrow  & \Downarrow \Omega R & \Downarrow \Omega R \nonumber\\
\left( \ba{c} n^i - m^i\\ - m^i
\ea \right)  \  \stackrel{Z_3}{\Longleftarrow} 
& 
\left(\ba{c}-n^i\\ m^i- n^i
\ea \right)   \ \stackrel{Z_3}{\Longleftarrow}   &
\left(\ba{c}m^i\\ n^i \ea \right)\nonumber\\
\label{array1}
\eeqa
We also denote the homology class of the i-th cycle of the a-th D6-brane as being 
defined as 
\beq
[ \Pi_a ] \ = \prod_{i=1}^3 (n_a^i \ [ a_i ]\  + \ m_a^i \ [ b_i ])  
\eeq  
We also denote by [$a_i$], [$b_i$] the basis for the
homology cycles across
the corresponding i-th two-tori lattice of the decomposable
six-dimensional
tori $T^6 = T^2 \times T^2 \times T^2$. 
The total homology class of the 06 planes is defined 
as
\beqa
[\Pi_{O6}] &=& [\Pi_{\Omega  R}] + [ \Pi_{\Omega  R  \omega} ]+ 
[\Pi_{\Omega  R  {\omega}^2 }  ] + [\Pi_{\Omega  R  {\theta} }]
+ [\Pi_{\Omega R \theta^2}]+  [\Pi_{\Omega R \theta^2 \omega}]
+  [\Pi_{\Omega R {\theta} \omega^2 }  ]+ \nonumber\\&&
[\Pi_{\Omega R \theta^2 {\omega}^2 }] + [\Pi_{\Omega R \theta {\omega} } ]
\eeqa
In turn the individual
homology classes of the cycles describing the action of spacetime and
worldsheet symmetries for the $\Omega  R$,  $\Omega  R  \theta$, $\Omega  R \omega $,  
 $\Omega  R  {\omega}^2$,
$\Omega R \theta^2$,   $\Omega R \theta^2 \omega$,
$\Omega R \theta {\omega}^2 $, $\Omega R \theta^2 {\omega}^2 $, 
$\Omega R \theta \omega$,
$\Omega  R  {\theta} \omega $,  actions on the 06-planes are given by 
\beqa
[ \Pi_{\Omega  R} ] = [ a_1 ] \times  [ a_2 ]
\times [ a_3 ], &    [ \Pi_{\Omega  R  \omega} ] = - [ a_1 ] \times
([ a_2 ]+[ b_2 ]) \times [ b_3 ],
\eeqa
\beqa
[\Pi_{\Omega  R  {\omega}^2 } ] = -[ a_1 ]
\times [ b_2 ] \times ([ a_3 ] + [b_3 ]), &  
[\Pi_{\Omega  R  \theta} ] = -([ a_1 ]+[ b_1 ])
\times [b_2] \times [ a_3 ],
\eeqa
\beqa
[\Pi_{\Omega R \theta^2}] = -[b_1 ] \times ([a_2 ]
+[b_2 ])
\times [a_3 ],&  [\Pi_{\Omega R \theta \omega}]
= -([a_1 ]+[b_1 ]) \times [a_2 ] \times[b_3],
\eeqa
\beqa
[\Pi_{\Omega R \theta^2 \omega}]
= [b_1 ] \times [b_2 ] \times
[b_3 ], & [\Pi_{\Omega R \theta^2 \omega^2}]
= -[b_1 ] \times [a_2 ] \times ([a_3 ]+[ b_3 ]),
\eeqa
\beqa
[\Pi_{\Omega R \theta {\omega}^2 }]= -([a_1 ]+[b_1 ])
\times ([a_2 ]+[b_2 ]) \times ([a_3 ]+[b_3 ])
\label{sxa1}
\eeqa
The O6-planes fixed under the orientifold actions $ \Omega  R$, ${\Omega  R  \omega}$, 
$\Omega  R  {\omega}^2 $, $ \Omega R \theta$,
$ \Omega R \theta^2$,$ \Omega R \theta {\omega}$, $ \Omega R \theta^2 \omega$, $\Omega R \theta^2 \omega^2$,
$\Omega R \theta {\omega}^2$ can be seen in figure (\ref{figo1}).
In $\Omega R $ orientifolds twisted crosscap tadpoles vanish 
\cite{twisted1, twisted2}, thus the 
orientifolds of ${\bf T^6/Z_3 \times Z_3}$ possess only untwisted RR tadpoles.
The images of a D6-brane under a $Z_3$ twist and the $\Omega R $
orientifold action may be denoted by
$[ \Pi_{a^{\prime}} ]$. 
The RR tadpole cancellation condition is equivalent to the vanishing of the 
RR charge in homology
\beq
\sum_a N_a \ [ \Pi_a ] \ +\ \sum_a
N_{a^{\prime}}\ [ \Pi_{a^{\prime}} ]\ +\ (-4)
\times [ \Pi_{O6} ] =\ 0 \ ,
\label{pa1}
\eeq
where $-4$ is the charge of the O6-plane.
Summing over the all the different homology classes the tadpole conditions
reduce to the general condition
\beq
\sum_a \ N_a \ { Z}_{[a]} =\ 4 \ ,
\label{tads}
\eeq
where ${Z}_{[a]} $ is given in appendix A.
A comment is in order. 
As all complex structure moduli are fixed, in N=1 supersymmetric constructions 
all NSNS tadpoles are absent. In non-supersymmetric constructions 
the only remaining NSNS disc tadpole is the one associated to the 
dilaton. However due to the absence of complex structure moduli in the $Z_3 \times Z_3$ 
constructions
the NSNS potential - which is of no 
interest to us in the present work - may exhibit the typical runaway dilaton potential behaviour
[see related comments for the $Z_3$ case \cite{lust3}].

\subsection{\small Massless spectrum}

The closed string sector has N=1 supersymmetry and the massless spectrum 
of vector and chiral 
multiplets can be found in \cite{twisted2}; 
the latter models are non-chiral as the 
D6-branes are parallel to the O6-planes. The models of the present work are chiral as the 
D6-branes wrap on general directions and are not parallel to the O6-planes. 
In the open string sector, 
we found that the net 
number of chiral fermions is given by the 
simple set of rules seen in table (\ref{matter}),
\begin{table}[htb] \footnotesize
\renewcommand{\arraystretch}{1.25}
\begin{center}
\begin{tabular}{|c|c|r|}
\hline\hline
\hspace{2cm} {\bf Sector} \hspace{1cm} &  Multiplicity
& {\bf Representation} \hspace{2cm} \\
\hline\hline
$a({\tilde \Theta}a)$  &       & $U(N_a)$ vector multiplet \hspace{3.3cm} \\
       & 3  & Adj. chiral multiplets  \hspace{3.3cm} \\
\hline\hline
$a({\tilde \Theta}b)$   & ${\tilde I}_{ab} $ & $(\fund_a,\antifund_b)$ fermions  \hspace{3.3cm} \\
\hline\hline
$a({\tilde \Theta}b')$ & ${\tilde I}_{ab^{\prime}}$ & $(\fund_a,\fund_b)$ fermions \hspace{3.3cm}  \\
\hline\hline
$a({\tilde \Theta}a^{\prime})$ & 
$\frac{1}{2} ( {\tilde I}_{aa^{\prime}} - \frac{1}{2} I_{a,O6} )$ &
$\Ysymm\;\;$     fermions \hspace{3.3cm}   \\
  & $\frac{1}{2} ({\tilde I}_{aa^{\prime}} + \frac{1}{2} I_{a,O6})$ &
$\Yasymm\;\;$   fermions  \hspace{3.3cm} \\
\hline\hline
\end{tabular}
\end{center}
\caption{\small General spectrum on D6-branes at generic angles
(namely, not parallel to any O6-plane in all three tori. 
The models contain additional non-chiral pieces
in the $aa'$, $ab$, $ab'$ sectors with zero intersection, if
the relevant branes have an overlap. 
\label{matter} }
\end{table}
where we denote as  
\beq
{\tilde I}_{ab}= \sum_{{\tilde \Theta}} I_{a({\tilde \Theta}b)}
\eeq
the intersection number (IN) from open strings stretching between the D6-brane $a$ and 
the D6-branes
taking values in the set of orbifold orbit elements 
\beq
{\tilde \Theta} = \{1,  \theta,  \theta^2, \omega,  \omega^2,  \theta \omega,
\theta^2 \omega, \theta \omega^2, \theta^2 \omega^2\}
\eeq 
Similarly the IN's coming from open strings stretching between the brane a and 
the images of the orbit for brane b are denoted as 
\beq
{\tilde I}_{ab^{\prime}}= \sum_{{\tilde \Theta}} I_{a({\tilde \Theta}b^{\prime})}
\eeq
In particular the intersection numbers between a brane $a$ and the orientifold images 
of the D6$_b$ brane orbits can be expressed in closed from as 
\beqa 
{\tilde I}_{ab} &=& 3({Z}_{[a]} {Y}_{[b]} - 
{Y}_{[a]} {Z}_{[b]})\nonumber\\
{\tilde I}_{ab^{\prime}} &= &3({Z}_{[a]}{Z}_{[b]}-
{Z}_{[a]}{Y}_{[b]} -{Y}_{[a]} {Z}_{[b]})
\nonumber\\
\# A &= & 3({Z}_{[a]} - 2 {Y}_{[a]})\nonumber\\
\# (A + S) &= & \frac{3}{2}({Z}_{[a]} -  2{Y}_{[a]})({Z}_{[a]}-1)
\label{rules}
\eeqa
in terms of the effective wrapping numbers ${Z}$, ${Y}$ 
defined in appendices A;B for the AAA; AAB, BBB lattices respectively.

The presence of spectrum rules is independent of the presence of 
any sypersymmetry that might be exhibited by the spectrum.    
N=1 supersymmetry may be preserved by a system of branes if each stack of 
D6-branes is related to the $O6$-planes by a rotation in $SU(3)$, that
is the angles ${\tilde \theta}_i $ of the D6-branes with respect to 
the horizontal
direction in the i-th two-torus obey the condition ${\tilde \theta}_1 + 
{\tilde\theta}_2 + {\tilde \theta}_3 = 0$. 
The supersymmetry of the 
models that is preserved by any pair of branes is determined by the 
choice of the orbifold and 
orientifold action. The models that are presented in this work do not have any 
supersymmetry preserved by a set of branes. 
In addition to the above chiral matter arising from bifundamentals  
there is also \footnote{and also non-chiral matter coming from
bifundamentals for which we comment on the end of section 4}
non-chiral massless matter present in the adjoint (NCMA).
NCMA arises from open strings stretching within a D$6_a$ brane and between a D$6_a$ brane and
its orbifold images as 
\beq
(Adj)_L :  \prod_{i=1}^3 \ ( (m_a^I)^2 + (n_a^I)^2 - m_a^I n_a^I)^2, \ \ A-lattice
\label{NCM1}
\eeq
\beq
(Adj)_L :  \prod_{i=1}^3 \ ( (m_a^I)^2 + 3 (n_a^I)^2 + 3 m_a^I n_a^I)^2, \ \ B-lattice,
\label{NCM2}
\eeq
with the non-zero contributions arriving from the $\theta^2 \omega$, $\theta \omega^2$ images. 
This sector has N=1 supersymmetry as the $Z_3 \times Z_3$ twist preserves 
it (see comments in the end of section 4).

\section{\large U(1) anomaly cancellation}

In any physical theory the existence of chiral 
fermions induces gauge anomalies which may be absent for the one loop
consistency of the theory. In the context of intersecting branes anomalies may 
cancel by the use of a generalized Green-Schwarz mechanism that couples the U(1) gauge 
fields $F_a$ to the untwisted RR fields $B_a$. This mechanism differs from the
corresponding mechanism in general type IIB orientifolds where the U(1) anomalies are 
cancelled through exchange of closed string RR twisted moduli [see also the old work 
in six- \cite{sagnot1} or 
four-dimensions \cite{ura}, \cite{bach}, \cite{sero}.
As it was pointed out in \cite{louis2} any U(1) gauge 
field that has a non-zero $B \wedge F$ coupling necessarily gets massive with a mass of the order 
of the string scale. In the low energy theory the broken symmetry remains as a global symmetry    
[For related issues in a general orientifold, with ``orthogonal branes'', see e.g.
\cite{kiri}, \cite{anasta}]. In fact,  
for the present $Z_3 \times Z_3$ orientifolds the anomaly cancellation has to take into account
the orbits (\ref{array1}) that the D6-branes wrap.
A sketch of the anomaly argument cancellation 
proceed as follows. 
The cubic non-abelian
SU(N$_a$) gauge anomaly (GGG) is actually the condition \cite{ibaba}
\beq
\sum_a N_b \ I_{ab} \ = \ 0
\eeq
which in the present constructions is proportional to
\beq
-2 {Y}_a \left(\sum_{b \neq a} N_b {Z}_b   \right) +
\sum_b N_b {Z}_a  {Z}_b + 
(N_a -4) ({Z}_a -2 {Y}_a) + 2 N_a ({Z}_a -2 {Y}_a)
({Z}_a -1)
\label{anoma1}
\eeq 
Eqn. (\ref{anoma1}) vanishes by the use of the RR tadpole condition 
in the first term
in (\ref{anoma1}).

The mixed U(1) anomalies should also cancel. In order to cancel 
the mixed U(1) gravitational  $ U(1)- g^2_{\mu \nu}$ anomalies that are 
proportional to
\beq
3 N_a ({Z}_a -2 {Y}_a)
\label{grav}
\eeq
and the mixed $U(1)_a - U(1)^2_b$ anomalies that are proportional to
\beq
N_a N_b({Z}_a -2 {Y}_a) Z_b
\label{mixed}
\eeq
we need to make use of a 
generalized Green-Schwarz mechanism that makes use of the 
mediation of the RR partners of the closed string untwisted geometric 
moduli.

In order to show that the various U(1)-anomalies cancel we will not 
use the usual picture with the D6-branes intersecting 
at angles but rather their T-dual picture where the D9-branes have magnetic 
fluxes along the six dimensional orbifolded tori of type I. In ten 
dimensions there are two RR fields, $C_2$ and $C_6$ with worldvolume 
couplings
\beq
\int_{D9_a} C_2 \wedge  F_a \wedge  F_a\wedge  F_a\wedge  F_a , \ \ 
\int_{D9_a} C_6 \wedge  F_a \wedge  F_a ,  \  I=1,2,3 \ .
\label{cole1}
\eeq
After dimensional reduction and taking into account the orbifold 
symmetry, (\ref{cole1}) reduces \footnote{ for the AAA torus} to the following 
Chern-Simons terms in the effective action for the D6-branes
\beqa
-6N_a ( {Z}_a -2 {Y}_a   ) \int_{M_4} B_2^0 \wedge F_a , &  \ 
0 \cdot  \int_{M_4} C_0^0 \wedge F_a \wedge F_a; \ 
\nonumber\\
-3 N_a ({Z}_a -2 {Y}_a ) \int_{M_4} B_2^I \wedge F_a , & \ 
-3{Z}_b \int_{M_4} C_0^I \wedge F_b \wedge F_b ,
\label{asd1}
\eeqa
where the 2-forms, $dC^0 = -^{\star} dB_2^0$, $dC^I =-^{\star}dB_2^I$ are defined as
\beq
B_2^0 = C_2, \ \  \  B_2^I = \int_{(T^2)^J \otimes (T^2)^K} C_6
\eeq
and their duals as 
\beq
C^I = \int_{(T^2)^I} C_2, \ \  C^0 = \int_{(T^2)^I \otimes (T^2)^J \otimes (T^2)^K} C_6
\eeq
These couplings have exactly the form required to 
cancel the mixed U(1) and gravitational anomalies (\ref{mixed}) 
and (\ref{grav}).

In general U(1) gauge fields that have a non-zero coupling to RR fields get massive  
while the associated U(1) survives as a global symmetry to low energies. 
From the form of the RR couplings to the U(1) gauge fields 
we derive that the only U(1) that becomes 
massive is the one given by the
expression
\beq
\sum_a \ N_a ({Z}_a -2 {Y}_a) \  F_a  
\label{massi}
\eeq 
All other U(1)'s that may be found - in a model building 
construction - may survive massless below the string scale and unless some 
Higgs mechanism \footnote{Such Higgs mechanisms have been employed in the construction of 
deformations of the 4-stack intersecting D6-brane toroidal orientifolds 
SM of \cite{louis2} in \cite{kokos5} and  \cite{kokos6}.}
 is involved that may give masses to them, 
they will also survive massless to low energies.

\section{\large  Non-supersymmetric D6-brane models with only the SM at low
energy}

In this section (and the following ones), we will use wrappings
 that have at least one zero electric or magnetic entry among them. As a result the 
models we construct may be non-supersymmetric.

Let us make the choice of wrapping numbers
\beq
(Z_a, Y_a) = \left( \ba{cc} 1,  & 1 \\\ea \right), \ (Z_b, Y_b) = 
\left( \ba{cc} 1, & 1 \\\ea \right), \ (Z_c, Y_c) = \left( \ba{cc}-1, & 0 \\\ea \right)
\label{wrap3}
\eeq 
where our three stacks assume the numbers 
$N_a =3$, $N_b = 2$, $N_c =1$.  
The RR tadpoles are satisfied and the spectrum can be seen in 
table (\ref{tabold}). The initial gauge group is a
$U(3)_a \times U(2)_b \times U(1)_c$. These models belong to the A$^{\prime}$-class and have 
no-exotics present.

\begin{table}[htb] \footnotesize
\renewcommand{\arraystretch}{2}
\begin{center}
\begin{tabular}{|r|c|c|}
\hline\hline
 ${\bf Matter\hspace{2cm}}$ & $(SU(3) \times SU(2))_{(Q_a, Q_b, Q_c)} \hspace{2cm}$ & $U(1)^{Y}$ \\
\hline\hline
$\hspace{2cm} \{ Q_L \} \hspace{2cm}$   & $3({\bar 3}, {\bar 2})_{(-1,\ -1,\ 0)} 
\hspace{2cm}$ & $1/6$  \\
\hline
$\hspace{2cm}\{ u_L^c \}\hspace{2cm}$   & $3(3, 1)_{(-2,\ 0,\ 0)}\hspace{2cm}$ & $-2/3$  \\
\hline
$\hspace{2cm}\{ d_L^c \}\hspace{2cm}$ & $3(3, 1)_{(1,\ 0, \  -1)}\hspace{2cm}$ & $1/3$ \\
\hline
$\hspace{2cm}\{ L \}\hspace{2cm}$ & $3(1, 2)_{(0,\  1,\  -1 )}\hspace{2cm}$ & $-1/2$ \\
\hline
$\hspace{2cm}\{ e_L^{+} \}\hspace{2cm}$ & $3(1, 1)_{(0, \ -2,\ 0)}\hspace{2cm}$ & $1$ \\
\hline
$\hspace{2cm}\{ N_R \}\hspace{2cm}$ & $3(1, 1)_{(0, \ 0, \ 2 )}\hspace{2cm}$ & $0$ \\
\hline
\end{tabular}
\end{center}
\caption{\small
A three generation SM chiral open string spectrum (A$^{\prime}$-model class). 
The required scalar Higgses may come from 
 bifundamental N=2 hypermultiplets in the N=2 $bc$, $bc^{\star}$ 
sectors \cite{louis2, kokos5, kokos6} that may trigger brane recombination.
\label{tabold}}
\end{table}
From the three initial U(1)'s one is anomalous and becomes massive by the GS 
mechanism by having a non-zero couplings to the RR fields, namely the 
\beq
U(1)^{massive} = 3 F_a + 2 F_b + F_c \ .
\eeq 
In addition there are also two anomaly free U(1)'s that correspond to the 
hypercharge and an extra U(1)
\beqa
U(1)^Y = \frac{1}{3}F_a - \frac{1}{2}F_b , \   \  \ 
U(1)^{ex}= \frac{3}{2} F_a 
+ F_b -\frac{13}{2}F_c 
\label{free}
    \eeqa
Hence we have found - table (\ref{tabold}) - exactly the chiral spectrum of 
the SM as at this point the spectrum for generic angles is 
non-supersymmetric. 
The same non-supersymmetric 
chiral spectrum construction was found in \cite{lust3} from
intersecting D6-branes in $Z_3$ orientifolds.
The Higgs fields that participate in the Yukawa couplings give masses to all
the chiral fields but the up quarks.

\begin{table}[htb] \footnotesize
\renewcommand{\arraystretch}{2}
\begin{center}
\begin{tabular}{|c|c|}
\hline\hline
 $Brane$ & $\hspace{2cm}(n^1, m^1) \times (n^2, m^2) 
\times (n^3, m^3)$\hspace{2cm} \\
\hline\hline
$\{ a \}$   & $\hspace{2cm}(0, \ 1) \times (1,\ 0)  \times (0,\ 1)\hspace{2cm}$\\
\hline 
$\{ b \}$   & $\hspace{2cm}(0,\ 1)  \times (1,\ 0) \times (0, 1)\hspace{2cm}$  \\
\hline
$\{ c \}$ & $\hspace{2cm}(0,\ \ -1) \times (0,\  1)  \times (-1,\ \ 0)\hspace{2cm}$ \\
\hline
\hline
\end{tabular}
\end{center}
\caption{\small 
Wrapping numbers responsible for the generation of the three stack 
D6-brane non-supersymmetric Standard Models of table (\ref{tabold}). 
\label{wrapthree} }
\end{table}

Let us now make a deformation of the previous choice 
of wrapping numbers,
\beq
(Z_a, Y_a) = \left( \ba{cc} 1,  & 1 \\\ea \right), \ (Z_b, Y_b) = 
\left( \ba{cc} 1, & 1 \\\ea \right), \ (Z_c, Y_c) = \left( \ba{cc}-1, & -1 \\\ea \right)
\label{wrap33}
\eeq 
It satisfies the RR tadpoles and 
corresponds to the spectrum seen in 
table (\ref{tabold}) but with reversed $U(1)_c$ charges. 
The exchange of the effective wrappings  
\beq
(Z, Y)_a \leftrightarrow (Z, Y)_b
\eeq
is a symmetry of the spectrum as the spectrum do not change.\newline
$\bullet$ {\bf Gauge couplings} \newline
The gauge coupling constants are controlled by the length of the corresponding cycles 
that the D6-branes wrap 
\beq
\frac{1}{\alpha_a} \ = \ \frac{M_s}{g_s}||l_i|| \ ,
\eeq  
where $||l_i||$ is the length
of the corresponding cycle for the i-th set of brane stacks. 
The canonically normalized U(1)'s as well the normalization of the abelian generators are given by
${\tilde U}(1)_a \ = \  \frac{F_a}{\sqrt{2 N_a}}$, $Tr(T_a T_b) = \frac{1}{2} \delta_{ab}$.
As the hypercharge is given \footnote{we used the conventions used in \cite{antotoma}} as a 
linear combination $Y(1)^Y = \sum_i c_i F_i$, the value of the weak angle becomes      
\beq
\sin^2 \theta_W = \frac{1}{1 + 4c_2^2 + 6c_3^2 (\alpha_2/\alpha_3)}  
\eeq
Taking into account that in the present models $\alpha_2 = \alpha_3$ we get 
the successful GUT relation
\beq
\sin^2 \theta_W \stackrel{M_s}{=} \frac{3}{8}
\label{mark}
\eeq
which means that the strong and the weak couplings unify at the unification 
scale, the string scale\footnote{Similar effects has been observed in \cite{gauge}.}.
Of course the really important issue here is weather or not partial unification can help us to 
confirm the experimental measured quantities like $sin^2 \theta_W$ and $a_{EM}$ at low energies.
These issues will be examined elsewhere. 
A comment is in order. In the models of this work, apart from the Standard model 
chiral matter we have also 
 present non-chiral bifundamental matter (NCBM) and also non-chiral adjoint matter (NCAM). 
The former arises in sectors that the D6-branes are parallel in at least one torus, 
the latter from sectors formed from open strings stretching between the 
D6-branes and their orbifold images [rules (\ref{NCM1}), (\ref{NCM2})]. 
In general NCAM, fermions and scalars, is believed 
to get massive - receiving radiative corrections - once supersymmetry is broken by 
massive N=1 supermultiplets running in 
loops by a mechanism that at present has been shown to be at work 
only \cite{louis2} at the level of the effective theory. 
If this mechanism is not at work at the level of string perturbation 
theory then the presence of extra NCAM - may destroy the
asymptotic freedom of the gauge groups and the result (\ref{mark}) will be useless.
Non-chiral matter (NCM) has only been shown that it gets massive in the context of 
Scherk-Schwarz 
deformations (SSD)  \cite{angel1} in toroidal orientifolds (TO); where only a subset 
of wrappings from the 
ones existing in the usual TO's gives masses to NCM. It remains to be seen if 
similar results also hold, once SSD is applied to the current orbifolds. We plan to return 
to this issue elsewhere.


\section{\large Non-susy models with the fermion spectrum of the N=1 SM and 
massive non-chiral exotics}

In this section we will generate a three stack non-supersymmetric model (B$^{\prime}$-model class) that 
generates the 
massless fermionic spectrum of the N=1 Standard Model at the string scale. Eventually, 
 all exotics present become massive due to the existence of appropriate Yukawa 
couplings. The only other known examples of models in the context of intersecting 
branes where all exotics become massive are the Pati-Salam $SU(4)_c \times SU(2)_L \times SU(2)_R$ 
GUTS of \cite{kokos1}. In the latter case, even though the GUT models are non-supersymmetric 
they do possess N=1 supersymmetric sectors, the latter being responsible for the
generation of gauge singlets.   

 The spectrum of the models can be 
seen in table (\ref{give11}). 
This model possess the SM chiral spectrum as well two massless Higgsinos and a pair of exotic 
triplets at the string scale. The initial gauge group at the string scale is a 
$U(3)_c \times U(2)_b \times U(1)_c$ which decomposes to an $SU(3)_c \times SU(2) \times U(1)_a 
\times U(1)_b \times U(1)_c$.
Subsequently, via the Yukawas
\beq
\lambda_C C_1 C_2 {\tilde N}_R \ + \ \lambda_H H_u H_d {\tilde N}_R ,
\eeq
the Higgsinos and the exotic colour triplets receive Dirac masses through the vev of the
tachyonic superpartner of the neutrino, ${\tilde N}_R$. 
As it is explained in the next section Higgsinos form a Dirac pair with a mass that 
receives the exponential suppression of the worldsheet area involved in the Yukawa 
couplings. Its mass scale can be anywhere below the string 
scale. 
The effective wrappings have been chosen to be  
\beq
(Z_a, Y_a) = \left( \ba{cc} 1,  & 0 \\\ea \right), \ (Z_b, Y_b) = 
\left( \ba{cc} 1, & 0 \\\ea \right), \ (Z_c, Y_c) = \left( \ba{cc}-1, & 0 \\\ea \right)
\label{wrap32}
\eeq 
At the top of the table (\ref{give11}) we see the massless fermion spectrum of the 
N=1 SM.  The corresponding superpartners are part of the massive spectrum and remain `hidden' in the 
 intersection of each corresponding fermion, unless they become tachyonic by varying 
appropriately the distances between the branes [see \cite{split01} for some relevant examples].

\begin{table}[htb] \footnotesize
\renewcommand{\arraystretch}{1.8}
\begin{center}
\begin{tabular}{|r|c|c|c|}
\hline\hline
 ${\bf Matter\hspace{2cm}}$ & Intersection & $(SU(3) \times SU(2))_{(Q_a, Q_b, Q_c)} \hspace{2cm}$ & $U(1)^{Y}$ \\
\hline\hline
$\hspace{2cm} \{ Q_L \} \hspace{2cm}$   & $ab*$ & $3({3}, { 2})_{(1,\ 1,\ 0)} \hspace{2cm}$ & $1/6$  \\
\hline
$\hspace{2cm}\{ u_L^c  \}\hspace{2cm}$   & $A_a$   & $3(3, 1)_{(2,\ 0,\ 0)}\hspace{2cm}$ & $-2/3$  \\
\hline
$\hspace{2cm}\{ d_L^c \}\hspace{2cm}$ &  $ac*$   & $3(3, 1)_{(-1,\ 0, \  -1)}\hspace{2cm}$ & $1/3$ \\
\hline
$\hspace{2cm}\{ \ L \}\hspace{2cm}$ & $bc*$  & $3(1, 2)_{(0,\  -1,\  -1 )}\hspace{2cm}$ & $-1/2$ \\
\hline
$\hspace{2cm}\{ \ H_d \}\hspace{2cm}$ &bc*$ $  & $3(1, 2)_{(0,\  -1,\  -1 )}\hspace{2cm}$ & $-1/2$ \\
\hline
$\hspace{2cm}\{ H_u \}\hspace{2cm}$ & $bc$   & $3(1, {\bar 2})_{(0,\  1,\  -1 )}\hspace{2cm}$ & $1/2$ \\
\hline
$\hspace{2cm}\{ e_L^{+} \}\hspace{2cm}$ & $A_b $  & $3(1, 1)_{(0, \ 2,\ 0)}\hspace{2cm}$ & $1$ \\
\hline
$\hspace{2cm}\{ N_R \}\hspace{2cm}$ & $S_c $  & $9(1, 1)_{(0, \ 0, \ 2)}\hspace{2cm}$ & $0$ \\
\hline
\hline
$\hspace{2cm}\{ C_1 \}\hspace{2cm}$ & $ac $  & $3(3, 1)_{(1, \ 0, \ -1 )}\hspace{2cm}$ & $1/3$ \\
\hline 
$\hspace{2cm}\{ C_2 \}\hspace{2cm}$ & $ac* $  & $3({\bar 3}, 1)_{(-1, \ 0, \ -1 )}\hspace{2cm}$ & $-1/3$ \\
\hline
\hline
\end{tabular}
\end{center}
\caption{\small
A three generation 4D non-supersymmetric model (B$^{\prime}$-model class) with the chiral content of 
N=1 MSSM on top of the table, in addition to $N_R$'s. There are three pairs
of $H_u$, $H_d$ Higgsinos. Note that this model
mimics models coming from gauge mediation scenarios and possess 
$sin^2 (\theta) = 3/8$ at $M_s$.  
\label{give11}}
\end{table}

Via the use of the generalized Green-Schwarz mechanism of section 3, the extra beyond the 
hypercharge U(1)'s become massive leaving only the hyperharge massless to low energies. 
The model possess three U(1)'s of which one becomes massive, namely - $3F_a + 2F_b - 3F_c$  - 
via the use of GS mechanism of 
section 3. The two remaining $U(1)'s$ are the hypercharge 
\beq
U(1)^Y = -\frac{1}{3}F_a + \frac{1}{2}F_b 
\eeq
and the $U(1)^{extra} = 3 F_a + 2 F_b +(13/3)F_c$ which is broken by ${\tilde N}_R$.  
Thus at low energy only the SM spectrum remains.
The Weinberg angle at $M_s$ in these models is also $sin^2 \theta = 3/8$, as the volumes of the 
3-cycles associated to the 
SU(3) and SU(2) gauge couplings agree as in the D-brane inspired models of \cite{antotoma}.   
These models belong to the B$^{\prime}$-model class differing in respect of the A$^{\prime}$-model class as they 
have in addition of the SM chiral spectrum extra massive non-chiral exotics.
In the next section we will examine an extended B$^{\prime}$-model class that possess extra gauge 
massive fermions and Higgsinos.


\section{\large Standard Models in the presence of massive exotics and Partial 
Split SUSY scenario}

Next we examine the construction of new non-supersymmetric vacua that break to 
the SM at low energy by using five stacks of intersecting D6-branes. 

The original gauge group
is a $U(3)_c \times U(2)_w \times U(1)_c \times U(1)_d \times U(1)_e$. 
The RR tadpoles are 
satisfied by the choices of effective wrappings

\beqa
(Z_a, Y_a) = \left( \ba{cc} 1,  & 1 \\\ea \right), \ (Z_b, Y_b) = 
\left( \ba{cc} 1, & 1 \\\ea \right), \ (Z_c, Y_c) = \left( \ba{cc}-1, & -1 \\\ea \right)\nonumber\\
(Z_d, Y_d) = \left( \ba{cc} -1,  & -1 \\\ea \right), \ (Z_e, Y_e) = 
\left( \ba{cc} 1, & 1 \\\ea \right),
\label{wrap300}
\eeqa
 where the have chosen
$N_a =3 $, $N_b =2 $, $N_c =1$, $N_d =1$, $N_e =1$. 
 The massless chiral spectrum can be seen in table (\ref{spli1}). 
The generators $Q_{a}$, $Q_{b}$, $Q_{c}$, $Q_{d}$ and $Q_{e}$ refer to the 
U(1) factors within the $U(3)_c$, $U(2)_w$, $U(1)_c$, $U(1)_d$,  $U(1)_e$,
respectively. There is one massive U(1), namely $- 3 F_a - 2F_b + F_c + F_d + F_e$.
There are also four U(1)'s that survive massless the GS mechanism, including the 
hypercharge, 
\beqa
U(1)^{(1)} = F_c - F_d, &  U(1)^{(2)} = F_c + F_d - 2F_e, \\
U(1)^{(3)} = 18 F_a + 12 F_b + 26 F_c + 26 F_d + 26 F_e, & U(1)^Y \ = \ \frac{1}{3}F^{a} - 
\frac{1}{2}F^b \ . 
\eeqa
The $U(1)^{(i)}$, $i=1,2,3$ generators could be broken by the vevs of the previously massive scalar 
superpartners of the $S^{(1)}$,  $S^{(2)}$, $S^{(3)}$ fermions respectively that become tachyonic, 
leaving only the hypercharge massless to low energies. 
The fermion singlets $S^{(I)}$ get masses by their couplings to their scalar tachyonic 
spartners which can get a vev.

 \begin{table}
[htb] \footnotesize
\renewcommand{\arraystretch}{1.6}
\begin{center}
\begin{tabular}{|c|c|c|}
\hline\hline
\  \ \  \ ${\bf Matter\  \ for \  \ Y^1}$ & \ $Y^1$ 
\ & ${SU(3) \times SU(2)_L}_{( Q_{a},\  \ Q_{b},\ \ Q_c,\ \ Q_d,\ \ Q_e )}\hspace{2cm}$  \\
\hline\hline
$\{ Q_L \}$  & $\frac{1}{6}$ & $3({\bar 3},\ 2)_{(-1,\ \ -1, \ \ 0,\ \ 0,\ \ 0)}$\hspace{2cm}  \\
\hline
$\{ u_L^c \}$ &  $-\frac{2}{3}$  & $3({\bar 3},\ 1)_{(-2,\ 0,\ \ 0, \ \ 0, \ \ 0)}$\hspace{2cm}  \\
\hline
$\{ d^c_L \}$ & $ \frac{1}{3} $ & $3( 3,\ 1)_{(1,\ \ 0,\ \ 0, \ \ 1, \ \ 0)}$\hspace{2cm} \\
\hline
$\{ e_L^{+} \}$ & $1$ & $3(1, \ 1)_{(0, \ \ -2, \ \ 0,\ \ 0, \ \ 0)}$ \hspace{2cm} \\
\hline
$\{ L \}$ & $-\frac{1}{2}$ & $3(1, \ 2)_{(0, \ 1, \ \ 0,\ \ 1, \ \ 0)}$ \hspace{2cm} \\
\hline
$\{ H_u \}$ & $\frac{1}{2}$ & $3(1, \ 2)_{(0, \ -1, \ \ 0,\ \ 0, \ \ -1)}
$ \hspace{2cm} \\
\hline
$\{ H_d \}$ & $-\frac{1}{2}$ & $3(1, \ 2)_{(0, \ 1, \ \ 1,\ \ 0, \ \ 0)}$ \hspace{2cm} \\
\hline
$\{ S_4 \}$ & $0$ & $3(1,\ 1)_{(0, \ \ 0,\ \ 0, \ \  1, \ \ 1)}$ \hspace{2cm} \\
\hline\hline
$\{ S_1 \}$ & $0$ & $3(1,\ 1, \ 1 )_{(0, \ \ 0,\ \ -2, \ \ 0, \ \ 0)}$\hspace{2cm} \\
\hline
$\{ S_2 \}$ & $0$ & $3(3,\ 1, \ 1 )_{(0,\ \ 0, \ \ 0, \ \ -2,\ \ 0)}$\hspace{2cm} \\
\hline
$\{  S_3 \}$ & $0$   & $3(1,\ 1, \ 1)_{ (0,\ \ 0, \ \ 1, \ \ 0,\ \ 1) }$ \hspace{2cm} \\
\hline
$\{  S_5 \}$ & $0 $   & $3(1,\ 1, \ 1)_{ (0,\ \ 0, \ \ -1, \ \ -1, \ \ 0)}$ \hspace{2cm} \\
\hline\hline
$\{  X_1 \}$ & $\frac{1}{3} $   & $3(3,\ 1)_{(1, \ \ 0, \ \ 1, \ \ 0, \ \, 0)}$ \hspace{2cm} \\
\hline
$\{  X_2 \}$ & $-\frac{1}{3} $   & $3({\bar 3},\ 1)_{(-1,\ \ 0,\ \ 0,\  \ 0, \ \ -1)}$ \hspace{2cm} \\
\hline
\end{tabular}
\end{center}
\caption{\small 
The three generation SM - from a five stack $SU(3)_C \times SU(2)_L \times U(1)_c
 \times U(1)_d   \times U(1)_e$. 
On the top of the table (a different B$^{\prime}$-model class) the fermionic spectrum of the 
N=1 SM with
 three pairs of Higgsinos and right handed neutrinos. 
Either one of the gauge singlets $S_I$ could be 
identified as the one
associated with the right handed 
neutrino. The exotics triplet $X_1$, $X_2$ and the $H_u$, $H_d$ fermion pair receive 
Dirac masses.  
\label{spli1} }
\end{table}
 
The gauge singlet fermions $S_I$ could be regarded as extra 
generations of right handed neutrinos. A good choice of right handed neutrinos is to 
choose the 
singlets $S_4 = N_R$. In this case -   $\langle  H_u \rangle = \upsilon_u $ -  the Yukawa 
couplings even though they come from dimension five operators
\beq
\frac{1}{M_s}L \ N_R \ \langle  H_u \rangle \ \langle S_2 \rangle \sim  \upsilon_u \ 
 L \ N_R
\label{saa1}
\eeq
they deliver a tree level Dirac mass term for neutrinos.
Alternative choices of right handed neutrinos are unsatisfactory as they result in
 suppressed masses.
Such a typical mass term is as follows.
Let us identify the singlet $S^1$ 
with $N_R^1$. The following Yukawa is allowed
\beq
e^{-A} (L N_R^1)  {{\tilde H}}_u^2  {{\tilde H}}_d  {\tilde S}_1  
{\tilde S}_2  ({\tilde S}_3)^2  {\tilde S}_4 \sim e^{-A} \frac{1}{M_s^2} \upsilon_u^2 \upsilon_d 
\ (N_L N_R^1) 
\eeq  
where we have consider that the scalars vevs $\langle {\tilde S}^I \rangle = M_s$,
 $\langle  {{\tilde H}}_u \rangle =  \upsilon_u$,  $\langle  {{\tilde H}}_d \rangle =  \upsilon_d$.
By tilde we denote the tachyonic scalar `superpartners' of 
the $S_I$, ${H}_u$, ${H}_d$ fermion singlets and Higgsinos respectively. 
The scalars  ${{\tilde H}}_u$, ${{\tilde H}}_d$ take part in electroweak symmetry breaking 
\footnote{
An alternative - but not phenomenologically interesting - possibility might be that 
the $S^1$ ($S^I$) are not {\em neutrinos}.  
Thus a candidate mass term for $S^1$ fermions (similar terms exist for the rest of 
$S^I$ fermions) may be
\beq
(S_1)^2  {\tilde S}_3^2 {\tilde S}_1^2 ({\tilde H}_u {\tilde H}_d)^4 \sim 
e^{-A^{\prime}}\frac{1}{M_s^7}\upsilon_u^4 \upsilon_d^4 
(S_1)^2 \equiv \frac{e^{-A^{\prime}} }{M_s^7} \tan^2({\tilde \theta}) \upsilon_d^4 (S_1)^2; 
\ \tan({\tilde \theta}) =\frac{\upsilon_u}{\upsilon_d}
\eeq 
which is very suppressed.}.
We also note of the possibility that the neutrino eigenstate is made of a linear combination 
of the
$S^I$ singlets; in this case also the dominant contribution to their mass comes from 
(\ref{saa1}).    

$\bullet$ $\bullet$ {\bf Split Susy scenario for intersecting branes ?}

Non-supersymmetric models in type I compactifications could solve the gauge 
hierarchy problem
 as the string scale could be lowered to the TeV in consistently 
with gravity as the Planck scale can become large, as long as there are large extra 
dimensions
transverse to the branes \cite{anto}. The only known string non-susy models that 
realize directly the large extra dimension scenario (LEDS) \cite{anto} and break to 
only the SM at 
low energy - making use of D5 branes on $T^4 \times C/Z_N$ - have been considered in 
 \cite{D5}. In this case there are two transverse dimensions and the 
string scale could be lowered to the TeV.  
In the present $Z_3 \times Z_3$ models the D6-branes wrap the whole of the internal 
space and thus the LEDS scenario cannot be applied. 
On the other hand the split susy scenario (SSS) \cite{split1} has been conjectured 
as a signal for high 
energy susy breaking for LHC \cite{split2}. In this 
scenario the candidate model could solve (potentially) the gauge hierarchy problem in a theory 
that breaks supersymmetry at a high scale, also 
accompanied by gauge coupling unification and 
should also predict simultaneously light Higgsinos
and gauginos. As it has been emphasized in \cite{split01} (and also suggested 
in \cite{split0} in a different context) in models coming from 
intersecting branes, the 
SSS scenario should be modified ({\em partial split susy criteria}) with respect
of the gauginos and Higgsino mass scales. Thus gauginos which 
are massless at tree level receive loop corrections from massive N=1 supermultiplets 
running in the loops \cite{louis2} and thus receive string scale 
masses \footnote{Even though the field theoretical study of \cite{louis2} was exhibited for
non-susy models in toroidal orientifolds the result also hold for N=1 susy models coming from 
orbifolds of orientifolds, as in the 
N=1 susy case one has to additionally take into account the different orbifold orbits.}. 
Also Higgsinos could  
form Dirac pairs and their mass range could be anywhere from $M_Z$ to the string scale. 
Hence the present 
models even though are not split susy models could also provide signatures for LHC (neglecting
stability/cosmological constant problems for non-susy/susy models respectively). 
The modified criteria of the SSS
proposal will be also verified using the models with the wrappings (\ref{wrap300}) in this 
section.

In general in the context of intersecting branes the existence of split susy models 
could be examined in either 
a N=1 
supersymmetric construction [ For some related work see \cite{kane1}] or in
constructions that have 
local supersymmetry as the SM models of \cite{cre1} that have been generalized
with the inclusion of B-field in \cite{kokosusy} (In \cite{kokosusy}
we have also considered the maximal five and six stack SM generalizations of \cite{cre1}).  
The latter models even though they are 
non-supersymmetric they have explicit the presence of N=1 
susy locally (see also related work on \cite{kane2}). On the other hand the 
present models where N=1 supersymmetry does not 
make its appearance even though 
are not split susy 
models can still satisfy some of the SSS existence
criteria namely like gauge coupling unification (GCU) and still predict 
light Higgsinos $H_u$, $H_d$.
In the models of table (5)
GCU is achieved for only two of the three gauge 
couplings; the strong and the weak unify at a string scale where the Weinberg angle 
is $sin^2 \theta = 3/8$. Gauginos becomes massive with a mass of order $M_s$ as usual in 
intersecting brane models. The Yukawa couplings are given by : 
\newline
$\bullet$ {\bf  100 GeV $<$ Light Higgsinos/Exotic triplets $<$ ${\bf M_s}$}
\beqa
Y_{[table \ (\ref{spli1})]} = \lambda_{\nu} L N_R {\tilde H}_u {\tilde  S}_2/M_s \ + \
\lambda_{e} L e_L^{+} {\tilde H}_d  {\tilde S}_5/M_s \ + \
\lambda_{d} Q_L d^c_L {\tilde H}_d {\tilde S}_5/M_s \ + \nonumber\\
\lambda_{H} H_u H_d ({\tilde S}_1 {\tilde S}_3 )/M_s  \ + \
\lambda_{X} X_1 X_2 ({\tilde S}_1 {\tilde S}_3 )/M_s 
\label{tax1}
\eeqa
where the tachyon scalars 
${\tilde H}_u$, ${\tilde H}_d$ play the role of the Higgs fields needed for 
electroweak symmetry breaking.
The Higgsinos $H_u$, $H_d$ and the triplets $X_i$ form Dirac 
mass terms in (\ref{tax1}).
If the area involved (the usual suppression factor in intersecting branes) in the 
relevant Yukawa (in Planck units) 
\beq
\lambda_{X,S}  \sim e^{-A}
\eeq
is vanishing and assuming that the scalar 
singlets ${\tilde S}_1$, ${\tilde S}_3$ get - their 
natural scale - vevs of order $M_s$, 
  then the Higgsino/triplets can reach their 
maximum value of order of the string scale, since $\lambda_{X,S}  \approx e^{-A} = 1$
[The Higgsinos get a tree level mass if no massive colour triplets 
are present in the models of the previous section].
On the other hand different value of the areas involved can make certain that lower 
mass scales are reached for the Higgsino and the colour triplets $X_i$ Dirac masses. 
Hence Higgsino Dirac masses of 100, 500 and 1000 GeV are obtained for area values 
of $\approx$ 32.9, 31.3, 30.6 respectively.




\section{\large Three generation non-supersymmetric flipped SU(5) GUTS}

The methodology to construct three generation non-supersymmetric 
flipped SU(5) GUTS that break to the SM at low energy have been exhibited in \cite{axeflo}; 
where it was shown how one can properly
identify the electroweak and Higgs multiplets in the intersecting brane flipped 
SU(5) (and SU(5)) context. These models \cite{axeflo} are considered within the $Z_3$ 
orientifold constructions of \cite{lust3}. 

In the present constructions it is also possible to construct flipped SU(5) and SU(5) 
models which may break to {\em only the SM} at low energy. 
The minimal case that we consider involves the presence of two stacks of 
D6-branes at the string scale. In this case, we can choose the effective D6-brane 
wrappings to be $(Z_a, Y_a)$,  $(Z_b, Y_b)$,
and solve the 
RR tadpoles (\ref{tads})
by making the choices $Z_a = 1$, $Z_b = -1$, and also 
\beq
(Z_a, Y_a) \equiv \left( \ba{cc} 1,  & Y_a \\\ea \right), \ (Z_b, Y_b) \equiv 
\left( \ba{cc} -1, & Y_b \\\ea \right),
\label{wrap31}
\eeq 
that corresponds to a two stack model with an 
initial  gauge group $U(5)_a \times U(1)_b$.
Since the nature of the intersection numbers
gives chiral fermions on intersections that are multiples of three,
in the general case we can assume that the number of generations 
is 3n and then solve for the 3G case, $n=1$. Having 3n generations 
enforces us to choose 3n copies of ${\bar {10}}$ representations. Hence an appropriate 
choice to generate flipped SU(5) models will be 
\beqa
I_{51} = 3n,& I_{51^{\star}} =0,& \#(A)_a = -3n, \  
\#(\Ysymm )_b = 3n,
\label{solu}
\eeqa 
 Explicitly the intersection numbers are given by
\beq
I_{51} \stackrel{{\bf \#(5,1)}}{=}3(Y_b + Y_a), \ \ (A)_a = 3(1-2Y_a), \ \ (S)_b = 3(2Y_b - 1)
\label{bein1}
\eeq
with the solution of (\ref{solu}), (\ref{bein1}) to be  
\beq
(Y_a, Y_b) = (\frac{n+1}{2} , \frac{n-1}{2}) 
\eeq
Choosing the model to have 3 generations, we recover the 
spectrum of table (\ref{tabo1}).   
\begin{table}[htb] \footnotesize
\renewcommand{\arraystretch}{1.25}
\begin{center}
\begin{tabular}{|c|c|c|c|c|c|}
\hline\hline
 ${\bf Sector}$  & Multiplicity
& ${\bf Repr}$ & $Q_a$ & $Q_b$ & $Q^{fl}$ \\
\hline\hline
$\{ 51 \}$   & $3$ & $({\bf 5}, 1)$  &$1$ &$-1$ & $3$ \\
\hline\hline
${\bf (A)_b }$ & $3$ & $({\bf \bar{10}}, 1)$ &$-2$ &$0$ & $-1$ \\
\hline\hline
${\bf S}$ & $3$ & $({ 1}, 1)$   & $0$ & $2$ & $5 $\\
\hline\hline
\end{tabular}
\end{center}
\caption{\small
Three generation flipped SU(5) GUT models. The last column indicates the flipped U(1) charge.  
\label{tabo1}}
\end{table}
The application of the Green-Schwarz mechanism of section 3,
suggests that only the U(1) gauge boson
\beq
U(1)^{fl} = \frac{1}{2} U(1)_5 - \frac{5}{2}U(1)_b
\eeq 
survives massless the Green-Schwarz mechanism by not having a non-zero 
coupling to the RR fields.  It exactly corresponds to the U(1) generator of the flipped 
$SU(5) \times U(1)^{fl}$. 
The generation content is as usual 
\beq
F = 10_1 = (u, d, d^c, \nu^c), \ \ f = {\bar 5}_3 = (u^c, \nu, e), \ \ l^c =1_5 = e^c 
\eeq    
A consistent set of wrappings for the three generation flipped SU(5) model
is given in table (\ref{higta})
\begin{table}[htb] \footnotesize
\renewcommand{\arraystretch}{1.25}
\begin{center}
\begin{tabular}{|c|c|c||}
\hline\hline
 $Brane/Gauge group$ & $N_a$ 
& $(n_a^1, m_a^1)(n_a^2, m_a^2)(n_a^3, m_a^3)$ \\
 \hline
U(5) & $5$ &(0, 1)(0, 1)(1, 1)\nonumber\\
\hline
U(1)  & $1$ &(1, 0)(0, 1)(0, 1)\\
\hline\hline
\end{tabular}
\end{center}
\caption{\small
D6-brane wrapping numbers for the three family flipped SU(5)(and SU(5)) 
GUT. 
\label{higta}}
\end{table}
where $(Z_5, Y_5)=(1,1)$ and $(Z_1, Y_1)=(-1,0)$.
The $SU(3) \times SU(2) \times U(1)_Y$ models of section (4) can be reinterpreted 
as coming from adjoint breaking of the $U(5) \times U(1)$ models of this section. 
During this process the adjoint ${\bf 24}$ gets a non-vanishing vev, and thus the 
U(5) stack splits into two stacks accommodating the U(3) and U(2) gauge groups.    
The reverse process corresponds to moving the U(3) and U(2) factors on top of 
each other by tuning the adjoint ${\bf 24}$ to a vanishing vev.

\subsection{\small Higgs sector in the flipped SU(5) GUTS}

The flipped SU(5) GUT symmetry breaks to the SM one by the use of the
tachyonic Higgs excitations seen in table (\ref{higta0}).
\begin{table}[htb] \footnotesize
\renewcommand{\arraystretch}{1.25}
\begin{center}
\begin{tabular}{|c|c|c|c|c|c|c|}
\hline\hline
 ${\bf Sector}$ & Field     & ${\bf Repr}$ & $Q_a$ & $Q_b$ & $Q^{fl}$ \\
\hline\hline
$ {\bf (A)_a} $ &  $H_1$  &${\bf 10}$  &$2$ &$0$ & $1$ \\
\hline
$ {\bf (A)_a} $  & $H_2$   &${\bf \bar{10}}$  &$-2$ &$0$ & $-2$ \\
\hline\hline
\end{tabular}
\end{center}
\caption{\small
GUT breaking Higgses for flipped SU(5) classes of GUT models. 
\label{higta0}}
\end{table}
Electroweak Higgses may come also from open strings stretching between the 
orbits of branes $5$ and $1^{\star}$.
Their quantum numbers may be seen in Table \ref{tabele}.
\begin{table}[htb] \footnotesize
\renewcommand{\arraystretch}{1.25}
\begin{center}
\begin{tabular}{|c|c|c|c|c|c|}
\hline\hline
 ${\bf Sector}$  & Higgs
& ${\bf Repr}$ & $Q_a$ & $Q_b$ & $Q^{fl}$ \\
\hline\hline
$\{ 51^{\star} \}$   & $h_1$ & ${\bf 5}$  &$1$ &$1$ & $-2$ \\
\hline
$\{ 51^{\star} \}$   & $h_2$ & ${\bf \bar{5}}$  &$-1$ &$-1$ & $2$ \\
\hline\hline
\end{tabular}
\end{center}
\caption{\small
Electroweak Higgses for flipped SU(5) classes of GUT models. 
\label{tabele}}
\end{table}

\subsection{\small Proton decay and Mass generation in the flipped SU(5) GUTS}
$\bullet$ {\em Proton decay}

The study of proton decay amplitude, with reference to an SU(5) GUT as those coming 
from $Z_2 \times Z_2$ orientifolds \cite{cve} has been performed to \cite{igorwi}. 
Further studies of the disk amplitudes that contribute to the gauge mediated proton decay 
(GMPD) in a general flipped SU(5) GUT and also in general
SU(5) GUTS - with emphasis on the orientifolds of $Z_3$ orbifolds has been performed 
in \cite{axeflo}. 
As in the present $Z_3 \times Z_3$ orientifold flipped SU(5) GUTS baryon number is 
not a gauged symmetry GMPD operators do contribute to proton decay - these 
contributions are identical to those appearing in \cite{axeflo} - and thus the string 
scale should be higher than $10^{16}$ GeV in order to safeguard the stability of the 
proton. The string scale in the present constructions is naturally high as \footnote{  
there are no dimensions transverse to the D6-branes that could be made large and lower the
string scale} the 
D6-branes wrap the whole of the internal space. \newline
$\bullet$ {\em Quark, lepton masses}
\beq
Y = \lambda^{quark}_u F \cdot {\bar f} \cdot \langle h_2 \rangle \ + \
\lambda^{lepton} f \cdot l^c  \cdot \langle h_1 \rangle 
\eeq
$\bullet$ {\em  Neutrino masses}
\beq
\lambda^{(1)} \ F \cdot  {\bar f} \cdot \langle h_2 \rangle \ + \
\lambda^{(2)} \ (F \cdot H_2) (F \cdot H_2) /M_s
\eeq
where in the first line there are mass terms for the up-quarks and charge lepton masses;
in the second line a see-saw mass matrix for the neutrinos.
There are no tree mass terms for the down 
  quarks but this is not a severe 
problem as the magnitudes of their masses is small and it is possible that they 
could be generated by higher non-renormalizable terms.

\section{\large SU(5) GUT generation}

SU(5) models may be produced by using the flipped SU(5) construction 
of the previous sections.
We choose to break the massless U(1)  - that survives the GS mechanism - of 
flipped SU(5) by turning on a singlet tachyon scalar
field coming from the open strings stretching between the branes 
that support the orbit of the $U(1)_b$ brane. Then our model becomes an SU(5)
class of GUTS.  The breaking to the SM in
this case is achieved by the use of the adjoint ${\bf 24}$, part of
the N=1 Yang-Mills multiplet in the aa-sector that utilizes itself by splitting 
the U(5) stack into two  
stacks of U(3) and U(2) branes - with identical wrappings - away from each other. Such a 
process have been described in \cite{split01}.
Further details on the identification of GUT and electroweak Higgses can be found in 
\cite{axeflo}.  We note that the Weinberg angle in these SU(5) GUTS also receives the
value $sin^2 \theta = 3/8$. This can be proved along the same lines as the ones 
used in \cite{lust3} as in the present discussion we have reproduced the spectra of SU(5) 
GUTS of \cite{lust3} in the context of $Z_3 \times Z_3 $ orientifolds using equal volume 
cycles for the relevant gauge couplings. 

\section{\large Pati-Salam models}

A non-supersymmetric Pati-Salam three family model  
could be constructed [the chiral spectrum may be seen in table \ref{patisalam1}] using 
three stacks of branes and the choice 
\beq 
(Z_a, Y_a) \equiv \left( \ba{cc} 1,  & 0 \\\ea \right), \ (Z_b, Y_b) \equiv 
\left( \ba{cc} 1, & 0 \\\ea \right), (Z_c, Y_c) \equiv 
\left( \ba{cc} -1, & -1 \\\ea \right),
\label{wrap3000}
\eeq 
giving a gauge group $U(4)_c \times U(2)_L \times U(2)_R$. The chiral content of these PS 
models (e.g. with surplus 3-, 6-plet exotics) is similar to the observable group chiral content 
of the PS models of tables 3, 4 in the 1st ref. of \cite{cve1}. 
 
 \begin{table}
[htb] \footnotesize
\renewcommand{\arraystretch}{1.4}
\begin{center}
\begin{tabular}{|c|c|c|c|c|}
\hline\hline
\  \ \  \ ${\bf Matter\ }$ & ${SU(4)_c \times SU(2)_L \times SU(2)_R }$ & 
$Q_{c}$ & $Q_{L}$ &  $Q_R$ \\
\hline\hline
$\{ F_L \}$       & $3(4,\ 2, \ 1)$        & $1$  &  $1$  & $0$  \\\hline
$\{ F_R \}$       & $3({\bar 4},\ 1, \ 2)$ & $-1$ &  $0$  & $1$  \\\hline
$\{ \omega_L \}$  & $3(6,\ 1, \ 1)$        & $2$  &  $0$  & $0$  \\\hline
$\{ \chi_L \}$    & $3(1,\ {\bar 2}, \ 2)$ & $0$  &  $-1$ & $1$  \\\hline
$\{ \psi_L \}$    & $3(1,\ 1, \ 3)$        & $0$  &  $0$  & $-2$  \\\hline
$\{ P_0 \}$       & $3(1,\ 1, \ 1)$        & $0$  &  $0$  & $2$  \\\hline
$\{ P_1 \}$       & $3(1,\ 1, \ 1)$        & $0$  &  $0$  & $-2$  \\\hline
$\{ P_2 \}$       & $3(1,\ 1, \ 1)$        & $0$  &  $2$  & $0$  \\\hline
\hline
\end{tabular}
\end{center}
\caption{\small 
Chiral spectrum for a three generation PS-model. The U(1) charges belong to the
respecting U(n) gauge groups.
\label{patisalam1} }
\end{table}

The PS models of table (\ref{patisalam1}) can be further subjected to gauge symmetry breaking
by adjoint splitting of the D6-branes. 
Hence the gauge symmetry can be broken directly to the SM by splitting the $U(4)_c$ stack  
- into parallel but not overlapping stacks, namely $a_1$ and $a_2$, made from 3 and 1 
branes - and also by splitting the $U(2)_R$ stack into two stacks, namely $c$, $d$, made from 
parallel 1 branes.
Moving away the D6-branes in the $U(4)_c$, $U(2)_R$ stacks \footnote{The adjoint 
breaking is further explained in \cite{cve, cve1}. } corresponds to 
giving vevs to the appropriate scalars in the adjoints
of $SU(4)_c$, $U(2)_R$. The application of the Green-Schwarz mechanism makes massive only the
$U(1)^{mas}=3F_{a_1} + 2F_b + F_{a_2}+F_c +F_d $. From the rest of the U(1)'s, 
one is the
$U(1)^Y = (-1/3)F_{a_1}+(1/2)F_b$ SM hypercharge which remains 
massless while the rest 
three massless U(1)'s namely,
$U(1)^{(1)}=2 F_{a_2}-F_c -F_d $, $U(1)^{(2)}=F_c -F_d $, $U(1)^{(3)}=-18F_{a_1} - 12F_b + 
13(F_{a_2}+F_c +F_d) $ receive masses if the tachyonic superpartners of the 
singlets $S_1$, $S_3$, $S_4$ receive a 
vev respectively.
Within this procedure the Pati-Salam gauge group may be broken
to the SM. 
The resulting spectrum is that of table (\ref{adjspli1}). The rest of chiral fermions; namely  
the colour triplets $X_i$ could receive a Dirac mass term of order $M_s$ by the coupling 
\footnote{Models with identical fermion content as in table (\ref{adjspli1}) which are constructed 
by direct methods and not by adjoint breaking -  
where the string scale mass terms for the exotic colour 
triplets arises at tree level - have been studied in \cite{split01}. See model A of the 
latter work.}  $X_1 X_2 {\langle S_1^B \rangle}  {\langle S_5^B \rangle} /M_s$,
where $S_1^B$,  $S_5^B$ tachyon superpartners of the $S_1$, $S_5$ fermions.  
 The singlet fermions $S^I$ also could receive a mass,
e.g. the $S^1$ obtain a term $(S^1)^2 {\langle \Psi^B_1 \rangle}^2 {\langle S_3^B \rangle}^2$ of order $M_s$, where $S_3^B$, $\Psi^B_1$ the tachyonic superpartners of the $S_3$ fermion and the ones coming from 
the $a_2 c*$ \footnote{singlet fermions in the  $a_2 c*$ intersection are non-chiral; 
in fact $I_{a_2 c*}=1-1=0$, the non-zero 
contributions - where the D6-branes are non-parallel in all tori -  are coming from the orbits 
$\Omega R \omega$, $\Omega R \theta^2 \omega^2$ and we have used the wrapping numbers 
(1,0)(0,1)(0,-1) and (1,1)(-1,0)(-1,-1) for the $a_2$ and $c$-branes respectively.
The non-chiral singlet fermions  $(\Psi_1)_{(0,\ 1,\ 0,\ 1, \ 0)}$, $(\Psi_2)_{(0,\ -1,\ 0, \ -1, \ 
0)}$ could also receive 
non-zero masses. E.g. $(\Psi_1)$ receive contributions of string scale mass from the terms
$ (\Psi_1) (\Psi_1) {\langle S_1^B \rangle}^2  {\langle S_3^B \rangle}^2 $ and
$ (\Psi_1) (\Psi_1) {\langle \Psi_2^B \rangle}^2  $
while 
$(\Psi_2)$ could get an $M_s$ mass from the coupling $  (\Psi_2) (\Psi_2)  {\langle \Psi_1^B \rangle}^2$.} 
intersection respectively. Thus at low energies only the SM fermion content remains (with no mass terms for the 
up-quarks).

\begin{table}
[htb] \footnotesize
\renewcommand{\arraystretch}{1.3}
\begin{center}
\begin{tabular}{|c|c|c|c|c|c|c|c|}
\hline\hline
 ${\bf Matter }$ & $SU(3) \times SU(2)_L$ & 
$Q_{a_1}$ &  $Q_{a_2}$  & $Q_b$ & $Q_c$ & $Q_d$ & $Y$ \\
\hline\hline
$\{ Q_L \}$  & $3(3, 2)$ & $1$ & $0$ &  $1$ & $0$ & $0$  & $\frac{1}{6}$  \\
\hline
$\{ u_L^c \}$ &  $3(3, 1)$ & $2$ & $0$ &  $0$ & $0$ & $0$  & $-\frac{2}{3}$  \\
\hline
$\{ d^c_L \}$ &  $3({\bar 3}, 1)$ & $-1$ & $0$ &  $0$ & $1$ & $0$  & $ \frac{1}{3}$  \\
\hline
$\{ e_L^{+} \}$ &  $3(1, \ 1)$ & $0$ & $0$ &  $2$ & $0$ & $0$  & $1$  \\
\hline
$\{ L \}$ & $3(1, \ {\bar 2})$ & $0$ & $0$ &  $-1$ & $0$ & $1$  
& $-\frac{1}{2}$  \\
\hline
$\{ H_u \}$ &  $3(1,  \ 2)$ & $0$ & $1$ &  $1$ & $0$ & $0$  & $\frac{1}{2}$  \\
\hline
$\{ H_d \}$ & $3(3, \ 1)$ & $0$ & $0$ &  $-1$ & $1$ & $0$  & $-\frac{1}{2}$  \\
\hline
$\{ S_1 \}$ & $ 3(1,\ 1)$ & $0$ & $-1$ &  $0$ & $1$ & $0$  & $0$  \\
\hline
$\{ S_2 \}$ & $3(3,\ 1)$ & $0$ & $-1$ & $0$ & $0$ & $1$  & $0$  \\
\hline
$\{  S_3 \}$ & $3(1,\ 1)$ & $0$ & $0$ &  $0$ & $-2$ & $0$  & $0$  \\
\hline
$\{  S_4 \}$ & $3(1,\ 1)$ & $0$ & $0$ &  $0$ & $0$ & $-2$  & $0$  \\
\hline
$\{  S_5 \}$ & $3(1,\ 1)$ & $0$ & $0$ &  $0$ & $-1$ & $-1$  & $0$  \\
\hline\hline
$\{  X_1 \}$ & $3(3,\ 1)$ & $1$ & $1$ &  $0$ & $0$ & $0$  & $-\frac{1}{3}$  \\
\hline
$\{  X_2 \}$ & $3({\bar 3},\ 1)$ & $-1$ & $0$ &  $0$ & $0$ & $1$  
& $\frac{1}{3}$  \\
\hline
\end{tabular}
\end{center}
\caption{\small 
The three generation SM from the adjoint splitting of the PS model 
of  table (\ref{patisalam1}). 
\label{adjspli1} }
\end{table}

\section{\large Conclusions}

In this work, we have described the construction of 
$Z_3 \times Z_3$ orientifolds where the D6-branes intersect at angles and are 
not parallel
with the O6-planes.  The presence of 
O6-planes requires for the cancellation of anomalies in the four 
dimensional compactification of IIA, intersecting D6-branes that wrap the 
internal six dimensional toroidal space and also a generalized Green-Schwarz 
mechanism to cancel the mixed U(1) anomalies. In this context, we have described in full 
generality the localization of chiral matter that gets consistently localized between 
the D6-branes.  
The presence of chiral matter is independent of the presence or not of N=1 
supersymmetry in the open string spectrum of the theory.  

We focused on the construction of models which break to the SM at low energy, which 
have  equal SU(3), SU(2) gauge 
couplings so that the Weinberg angle at $M_S$ is equal to 3/8 \footnote{Problems 
related to the successful prediction through RG running of low energy phenomenological quantities 
like $sin^2 (M_z)$, $a_{EM}$ 
may be considered elsewhere.}.
We have also described the possibility of constructing a) GUT models and also b) new SM vacua (section 6)  with
the spectrum of the SM in addition to (3) pairs of Higgsinos and exotic triplets (the latter pairs becoming massive by tree level Yukawa couplings)
and c) the construction of different 
(wrapping) solutions to 3- (A$^{\prime}$ class) and 5-stack (B$^{\prime}$ class) 
non-supersymmetric models with only 
the SM at low energy than the ones extensively studied in
\cite{split01}. At 3-stacks we get only the SM without any chiral exotics 
present. These 3-stack models exactly reproduce the SM example given in \cite{lust3} in 
the context of $Z_3$ orientifolds \footnote{as we have already said in section 4 we 
neglected through this work the issue of massless non-chiral matter}.
\newline
At 5-stacks we get chiral models with the 
chiral fermionic spectrum of the N=1 supersymmetric SM in addition to three identical pairs 
of massive non-chiral exotics. Even though the models are non-supersymmetric they 
predict Higgsinos with a light mass that could provide us with a signal for LHC. 
The construction of GUT models is also possible; in the last section where we have detailed the 
construction of two stack 
flipped SU(5) (also SU(5)) models which can break to the
SM at low energy and also three stack Pati-Salam GUTS which after adjoint breaking break 
to the B$^{\prime}$-model class.

For the non-supersymmetric SM's, SU(5) and Pati-Salam 
$SU(4) \times SU(2)_L \times SU(2)_R$ GUTS - of the present work there is an
 absence of tree level mass term for the up-quarks.  This is a general problem in models 
involving antisymmetric representations [The same problem persists
also in the construction of N=1 SU(5) 
models from $Z_2 \times Z_2$ \cite{cve10} orientifolds and in the N=0 models 
of \cite{lust3}.].  
 Due to the largeness of the top quark mass, it will be
 impossible to generate a mass
from higher order non-renormalized corrections. 
On the contrary in flipped SU(5) GUTS, the opposite happens as there are no mass terms for the
d-quarks. As the d-masses are generally small, the absence of a tree level mass is not very 
problematic, as higher order terms may generate in principle the required masses. 

As the models we constructed are non-supersymmetric and there 
are no transverse dimensions that can dilute the strength of gravity, 
the string/GUT scale may be high \cite{anto}. Hence they can safeguard the models from 
proton decay since baryon number is not a gauged 
symmetry. However, the high scale  
is unsatisfactory since we want also to avoid the gauge hierarchy problem (GHP) in the 
Higgs sector.

In all the models that we have constructed in this work and well as in the
 preceding \cite{split01} one, we have constructed D-brane models using in all cases 
wrappings $(n^I, m^I)$ which have at least one zero entry for the AAA vacua.  
Since the models derived in 
all cases were non-supersymmetric to construct N=1 supersymmetric models
we might have to use wrappings with all entries being non-zero for 
the AAA tori \footnote{especially for the AAA tori such a procedure do not guarantee
the construction of N=1 vacua}.
Such an attempt requires 
an extensive computer search and will be the subject of future work. We also note that we 
have not 
investigated the construction of N=1 or N=0 models using the AAB, BBB tori for which we have 
presented 
the RR tadpoles, the spectrum rules and the Green-Schwarz anomaly cancellation mechanism 
in appendix B. We leave this task for future work.

The present constructions even though the orbifold symmetry stabilizes naturally 
the complex structure moduli do not fix K\"ahler moduli that are introduced by the 
orbifold in its twisted sectors. We note that more moduli could be fixed in 
models with
RR and NSNS fluxes e.g. \cite{flu1},\cite{flu2} but in this case the exact 
string description is lost since 
the results are valid in the low energy supergravity approximation. 
Alternatively, one could use 
oblique magnetic fluxes \cite{ma1},\cite{ma2} which can fix in principle all moduli  but where 
no gauge group factors of rank large enough to accommodate the SM arise at the moment. On the 
contrary in the present work we have chosen to build realistic gauge groups, leaving aside
problems related to stabilization, fixing moduli and gauge hierarchy that finally render our 
models semirealistic.      
Some of these moduli in semirealistic models from 
orbifolded orientifolds could be 
fixed in principle by the use of multiple gaugino condensates [see for example 
\cite{cve} for examples of this method in N=1 semirealistic models]. We also note that the 
construction of N=1 supersymmetric models - that have a high scale - may alleviate 
the hierarchy problem that the 
present models possess. It will also be interesting to see what will be the effect of discrete 
torsion in the present models \cite{cri1}.

 As the models we have constructed are non-supersymmetric
 only the NSNS dilaton tadpole  
remains uncancelled. The presence of the dilaton tadpole does not signal an 
inconsistency of the theory, but rather that the background is not a solution of the 
classical equations of motion[see some recent work \cite{newdi}] and thus may be 
corrected \cite{du, blu}. In fact it plays the role 
of an uncancelled effective cosmological constant and is rather connected 
with the problem of breaking supersymmetry. As even in N=1 supersymmetric models, where no 
NSNS tadpoles  
are present initially and after the breaking of supersymmetry we do get a cosmological 
constant of the wrong size, we will choose to ignore the NSNS issue for the time 
being as it is connected with whatever mechanism will solve the
 cosmological constant problem in particle physics.

\begin{center}
{\bf Acknowledgments}
\end{center}
I am grateful
to R. Blumenhagen and A. Uranga
for useful discussions. This work is supported 
via the programme 
``Pythagoras I'' Grant K.A. 70-3-7310 .

\section{\large Appendix A}

In this appendix we list the analytic form of the effective 
wrapping numbers, $Z_{[a]}$, $Y_{[a]}$ in terms of the `electric' and 
'magnetic wrapping numbers $(n_a^i, m_a^i)$.

$\bullet$ {\bf AAA torus}
\beqa
Z_{[a]}^{AAA}= (2 m_a^1 m_a^2 m_a^3 + 2 n_a^1 n_a^2 n_a^3 -n_a^1 m_a^2 m_a^3 -
m_a^1 n_a^2 m_a^3 - m_a^1 m_a^2 n_a^3 -n_a^1 n_a^2 m_a^3 &\nonumber\\ 
-n_a^1 m_a^2 n_a^3 -m_a^1 n_a^2 n_a^3) , &   \nonumber\\
Y_{[a]}^{AAA} =  (n_a^1 n_a^2 n_a^3 + m_a^1 m_a^2 m_a^3 -n_a^1 n_a^2 m_a^3
-n_a^1 m_a^2 n_a^3 -m_a^1 n_a^2 n_a^3)
\label{wrapa1}
\eeqa
We can introduce an abbreviation that can help us simplify and shorten 
the lengthy expressions. Alternatively we can write (\ref{wrapa1}) as
\beqa 
Z_{[a]}^{AAA} = (2 m m m + 2 n n n - \underline{n m m} 
 -\underline{n n m})_a &\nonumber\\ 
Y_{[a]}^{AAA} =  (n n n + m m m - \underline{n n m} )_a
\eeqa
where each triplet of letters correspond to - reading from 
left to right  - to the wrappings of the first, second and third tori 
respectively. A single letter inside a triplet of wrappings 
that may be different than the rest of them is always permuted among the different tori.

\section{\large Appendix B}

In order to derive the spectrum, RR tadpoles and apply the 
Green-Schwarz anomaly cancellation mechanism in those classes of 
$Z_3 \times Z_3 $ four dimensional orientifold string compactifications
that the internal space is made of BBB, AAB tori 
we may use a different notation for the A lattice than the one 
used in the main body of the paper. 
The notation for the A-, B- lattices we will follow - in this appendix - may be 
as in \cite{lust3}.
 In particular we will go
to a point in moduli space where the complex structure in all three $T^2$'s will 
be fixed to the values 
$U_A = \frac{1}{2} + i \frac{\sqrt3}{2}$, $  U_B = \frac{1}{2} + i \frac{1}{2\sqrt{3}}$.

Hence under the $Z_3$ and $\Omega R$ symmetries the brane orbits are given for the
A-torus by  
\beqa
\left( \ba{c} \n^i\\ \m^i
\ea \right)  \  \stackrel{Z_3}{\Longrightarrow} 
& 
\left(\ba{c}-n^i-m^i\\ n^i 
\ea \right)   \ \stackrel{Z_3}{\Longrightarrow}   &
\left(\ba{c}m^i\\ - n^i-m^i \ea \right)\nonumber\\
 \Omega R\Downarrow  & \Downarrow \Omega R & \Downarrow \Omega R \nonumber\\
\left( \ba{c} n^i + m^i\\ - m^i
\ea \right)  \  \stackrel{Z_3}{\Longleftarrow} 
& 
\left(\ba{c}-m^i\\ -n^i
\ea \right)   \ \stackrel{Z_3}{\Longleftarrow}   &
\left(\ba{c}-n^i\\ n^i+m^i \ea \right)\nonumber\\
\eeqa
and for the B-torus by 
\beqa
\left( \ba{c} n^i\\ m^i
\ea \right)  \  \stackrel{Z_3}{\Longrightarrow} 
& 
\left(\ba{c}-2n^i- m^i\\ 3n^i + m^i
\ea \right)   \ \stackrel{Z_3}{\Longrightarrow}   &
\left(\ba{c}n^i+ m^i\\ - 3n^i -2m^i\ea \right)\nonumber\\
 \Omega R\Downarrow  & \Downarrow \Omega R & \Downarrow \Omega R \nonumber\\
\left( \ba{c} n^i + m^i\\ - m^i
\ea \right)  \  \stackrel{Z_3}{\Longleftarrow} 
& 
\left(\ba{c}n^i\\ -3n^i- m^i
\ea \right)   \ \stackrel{Z_3}{\Longleftarrow}   &
\left(\ba{c}-2n^i-m^i\\ 3n^i + 2m^i \ea \right)\nonumber\\
\eeqa

$\bullet$ {\bf AAB torus}

The massless spectrum is given by the rules of eqn's (\ref{rules}).
The RR tadpoles are given by
\beq
\sum_a N_a \ Z_{[a]}^{AAB} = 4 \ ,
\eeq
where
\beqa
 Z_{[a]}^{AAB}&=& 2 n_a^1 n^2_a n^3_a + m_a^3 n_a^1 n_a^2 - m_a^1 m_a^2 n_a^3 +  
m_a^2 n_a^1 n_a^3 + 
m_a^1 n_a^2 n_a^3  - m_a^1 m_a^2 m_a^3 \nonumber\\
&=& (2 n^1 n^2 n^3 - m^1 m^2 n^3- m^1 m^2 m^3 +
\underline{mnn})_a
\eeqa
and
\beq
 Y_{[a]}^{AAB} =  - m_a^1 m_a^2 m_a^3 - n_a^1 m_a^2 m_a^3 - 
m_a^1 n_a^2 m_a^3 - 2m_a^1 m_a^2 n_a^3 - n_a^1 m_a^2 n_a^3 -  m_a^1 n_a^2 n_a^3 
+ n_a^1 n_a^2 n_a^3
\eeq


$\bullet$ {\bf BBB torus}

The massless spectrum is given by the rules of eqn's (\ref{rules}).
The RR tadpoles are given by
\beq
\sum_a N_a \ Z_{[a]}^{BBB} = 12 \ ,
\eeq
where
\beqa
 Z_{[a]}^{BBB}&=& 6n_a^1 n^2_a n^3_a + 3m_a^1 n_a^2 n_a^3 + 
3n_a^1 m_a^2 n_a^3 + 3n_a^1 n_a^2 m_a^3 + 
m_a^1 m_a^2 n_a^3 +  \nonumber\\
&&m_a^1 n_a^2 m_a^3 + n_a^1 m_a^2 m_a^3  \nonumber\\
&=& {(6 n^1 n^2 n^3 + 3 \underline{mnn} +\underline{mmn} )}_a  \nonumber\\
\eeqa
and
\beqa
 Y_{[a]}^{BBB} &=&  - m_a^1 m_a^2 m_a^3 + 3 n_a^1 n_a^1 n_a^1 - 
m_a^1 m_a^2 n_a^3 - m_a^1 n_a^2 m_a^3 - n_a^1 m_a^2 m_a^3 \nonumber\\ 
&=&   (-m^1 m^2 m^3 + 3 n^1 n^2 n^3 - \underline{mmn})_a
\eeqa

$\bullet \bullet$ {\bf U(1) anomaly cancellation}

The Green-Schwarz mechanism - for the BBB, AAB tori - 
makes massive the U(1) that has its couplings 
given by (\ref{massi}). Thus for the BBB vacua 
the examination of the BF couplings reveals e.g.
\beq
  -18 (Z_a - 2 Y_a )  
\int_{M_4} B_2^o \wedge F_a ,\  \ \ \  -6 Z_b   
   \int_{M_4}  C^o \wedge F^b \wedge F^b ;
\eeq  
that these couplings have the right form to cancel the mixed-U(1) gauge 
anomalies via the generalized Green-Schwarz mechanism  
\newpage

{

\end{document}